\documentclass[prl,twocolumn,showpacs,showkeys,preprintnumbers,amsmath,amssymb]{revtex4}

\usepackage{amsmath}    
\usepackage{graphicx}   
\usepackage{verbatim}   
\usepackage{subfigure}  
\usepackage{hyperref}   

\usepackage{dcolumn}
\usepackage{bm}

\begin{document}


\title{Percolation in a Class of Band Structured Random Matrices}
\author{Dieter~W.\ Heermann}
 \email{heermann@tphys.uni-heidelberg.de}
\author{Manfred Bohn}
\affiliation{Institut f\"ur Theoretische Physik, Universit\"at Heidelberg,
  Philosophenweg 19, D--69120 Heidelberg, Germany }

\date{\today}

\begin{abstract}
We define a class of random matrix ensembles that pertain to random looped polymers.
Such random looped polymers are a possible model for bio-polymers such as chromatin
in the cell nucleus.
It is shown that the distribution of the largest eigenvalue $\lambda_{\mbox{max}}$
depends on a percolation
transition in the entries of the random matrices. Below the percolation threshold the
distribution is multi-peaked and changes above the threshold to the Tracy-Widom
distribution. We also show that the distribution of the eigenvalues is neither of the Wigner form
nor gaussian.
\end{abstract}

\maketitle

Statistical properties of complex systems can be described by random matrix 
ensembles~\cite{reviews}. Most of the work has been concentrated on describing physical
systems where the resulting random matrix elements all are independent and identically
distributed (iid) or ensembles of matrices where the probability distribution is invariant
with respect to orthogonal or unitary transformations. 
These two possible generalizations
of the gaussian unitary (orthogonal) ensemble represent different dependencies of the entries.
While the first possibility has arbitrarily distributed entries, the second imposes strong statistical
dependence over long distances resulting in different properties of the 
eigenvalue spectrum~\cite{Wigner,Boutet-1995}. Common to the  random matrix ensembles 
based on the first possibility is the Wigner semicircle~\cite{Wigner}
distribution for the eigenvalue spectrum and the Tracy-Widom distribution~\cite{Tracy-Widom} for the largest
eigenvalue, even in the case of a band structured matrix~\cite{Casati-1991}. 

\begin{figure}
\begin{center}
\includegraphics[width=\columnwidth]{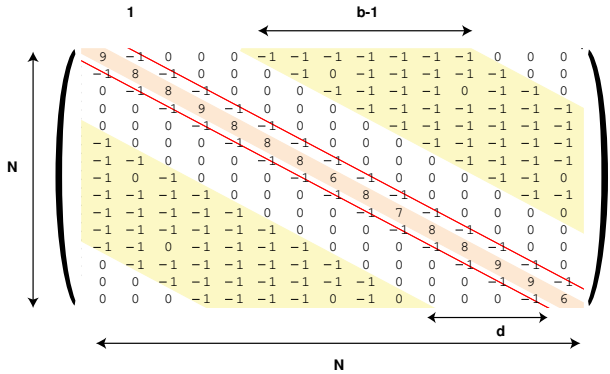}
\caption{\label{fig:matrix} Shown is the construction and a sample of the random matrices}
\end{center}
\end{figure}

For biological systems it turns out that random matrices are needed where all
of the matrix elements are iid except for the main diagonal and first off-diagonal~\cite{Bohn-2007}, 
to describe, for
example  the conformations and three-dimensional organization of
chromatin in the cell.  In some case
it is necessary to eliminate some matrix elements resulting in a band structured random
matrix. Band structure random matrices have also been considered for discretized models
of solid state physics~\cite{Wilkinson-1991}. 
Thus, while most entries are independent, the diagonal introduces
correlations. It is thus natural to suspect that the resulting eigenvalue spectrum may not
belong to either of the two universality classes. 
 
In this work we define an ensemble of band structured random matrices whose matrix elements
can take on two possible values with the main diagonal matrix elements as dependent random 
variables. We are interested in the resulting eigenvalue spectrum and the distribution of the largest 
eigenvalue in terms of a percolation transition. 
 
\begin{figure}
\begin{center}
\includegraphics[width=\columnwidth]{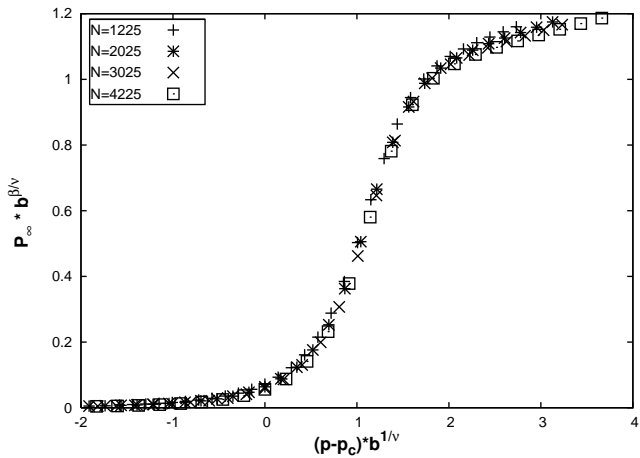}
\caption{\label{fig:p-infty}Finite-size scaling of the percolation probability for the ratio
of matrix size $N$ and band width $b$ of $b^2/N =1$.
The exact values of the critical exponents  for the two-dimensional site percolation
were used. The non-universal value for $p_c$ was determined to be $0.61$. Other
ratios $b^2/N = \mbox {const}$ also produced good scaling. 
The data presented in the figure  were obtained
by averaging over at least $10000$ random matrix configurations for each $p$ and matrix
size and band width combinations.}
\end{center}
\end{figure}

Let $b$ denote the bandwidth of a $N\times N$ matrix with a displacement $d$ from
the main diagonal ($N\ge b > 1$) with matrix elements (see Figure~\ref{fig:matrix})
\begin{equation}
H_N(i,j) = N^{-a}h(i,j) ,
\end{equation}
where 
\begin{equation}
H_N(i,j)=0 ~~\mbox{for}~~
\left \{ \begin{array}{l}
 |i-j-d|>b    \\
 |i-j|<d, i\ne j, j\ne i-1,i+1
  \end{array}\right . 
\end{equation}
and, within the band, we have Bernoulli variables
\begin{equation}
\begin{array}{ll}
 h(i,j) = -1 & \mbox{with probability} \; p\in [0,1]  \\
 h(i,j) = 0 & \mbox{with probability} \; 1-p  \\
 \end{array}
\end{equation}
with the additional constraints
\begin{equation}
\begin{array}{ll}
h(i-1,i)=h(i,i+1) = -1, &\quad 2\le i \le N-1 \\
h(i,i) = \sum_{|i-j-d| \le b} |h(i,j)| .
\end{array}
\end{equation}
$N^{-a}$ is a norming factor for the eigenvalues. Thus the matrices that we consider
are real symmetric band structured random matrices. The band consists of random
variables that take on the values $-1,0$ and the band may have been 
shifted. The main diagonal is the sum of the off-diagonal random values with the
exception of the first off-diagonal which is never shifted and has always values $-1$. 
Thus the matrix is diagonally dominant.

This structure mimics a polymer chain, where the first off-diagonal ensures the
integrity of the chain, the diagonal gives the coupling strength and the other off-diagonal
elements mimic random loops. Note that the regular polymer case of $b=1, d=0$ 
with gaussian random variables has been solved by Dyson~\cite{Dyson-1953}.

\begin{figure}
\begin{center}
\includegraphics[width=\columnwidth]{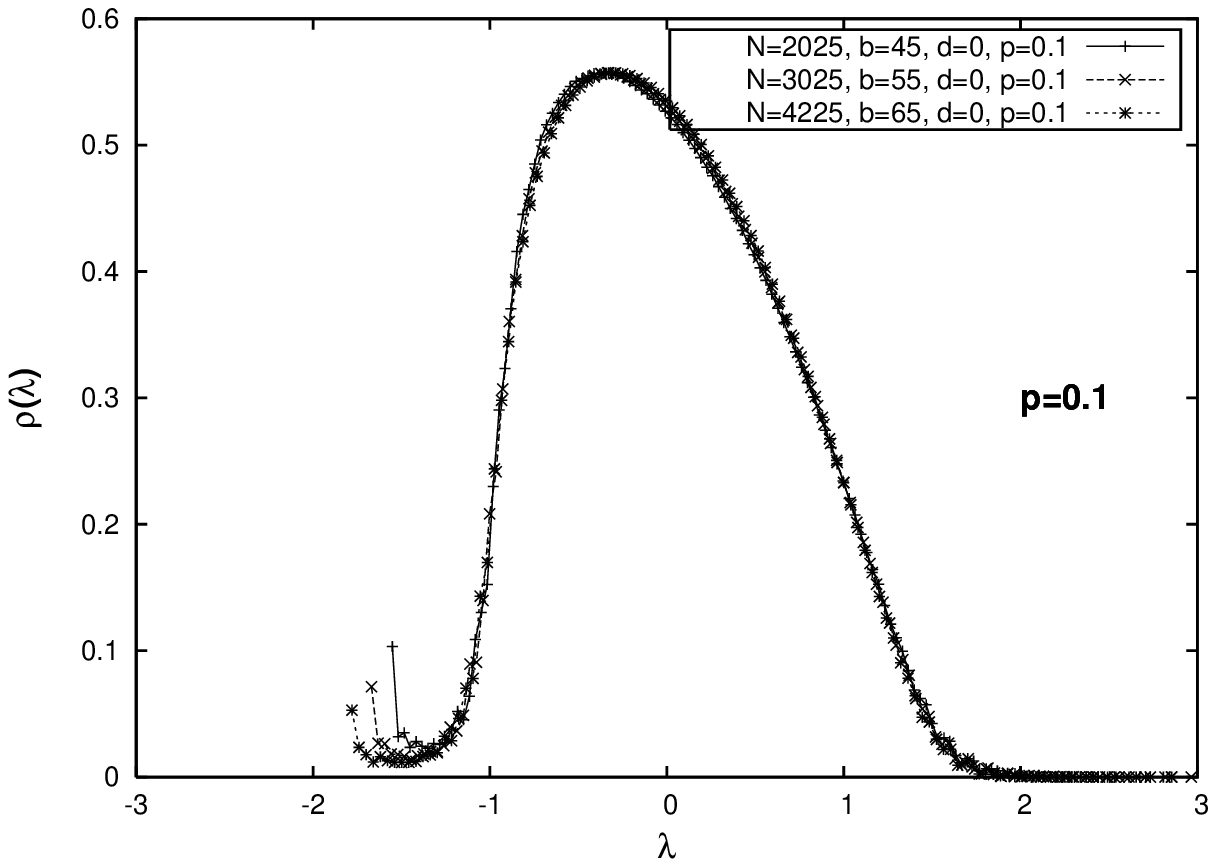}
\includegraphics[width=\columnwidth]{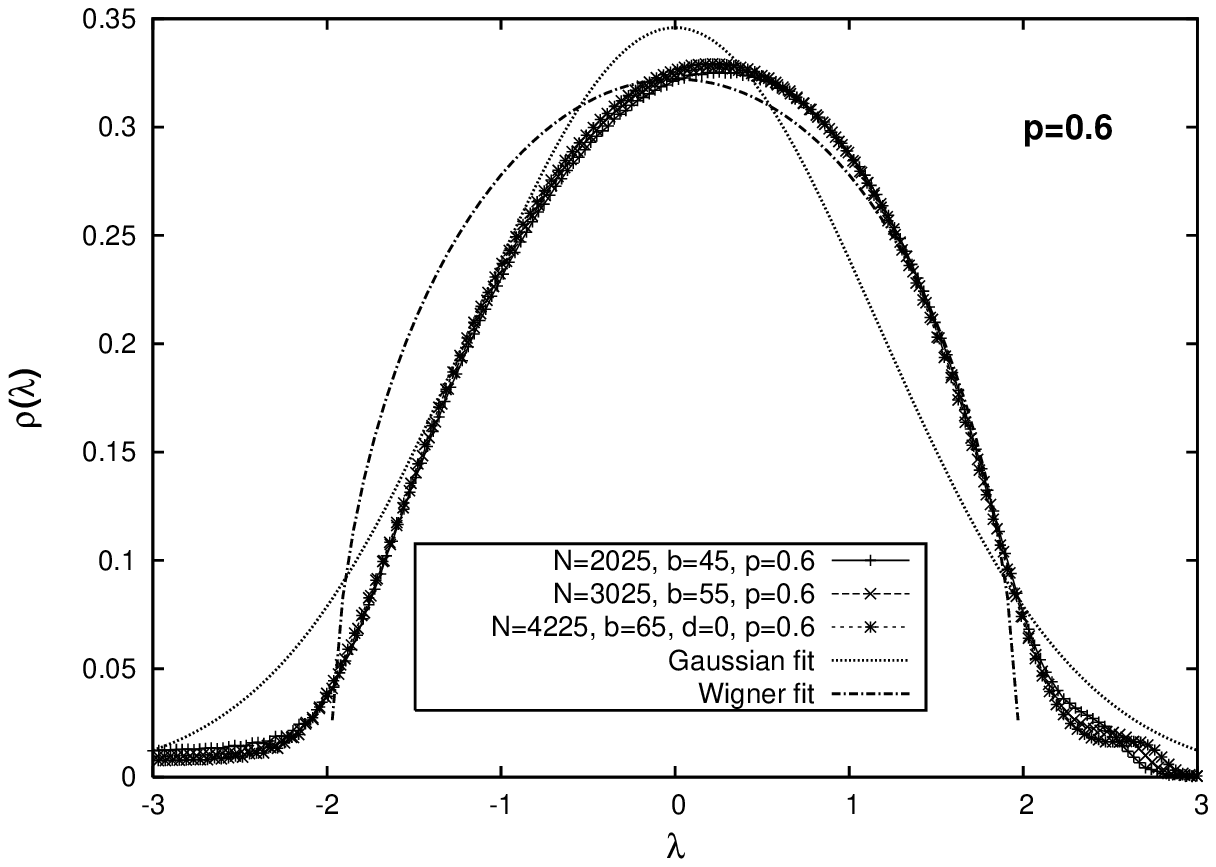}
\includegraphics[width=\columnwidth]{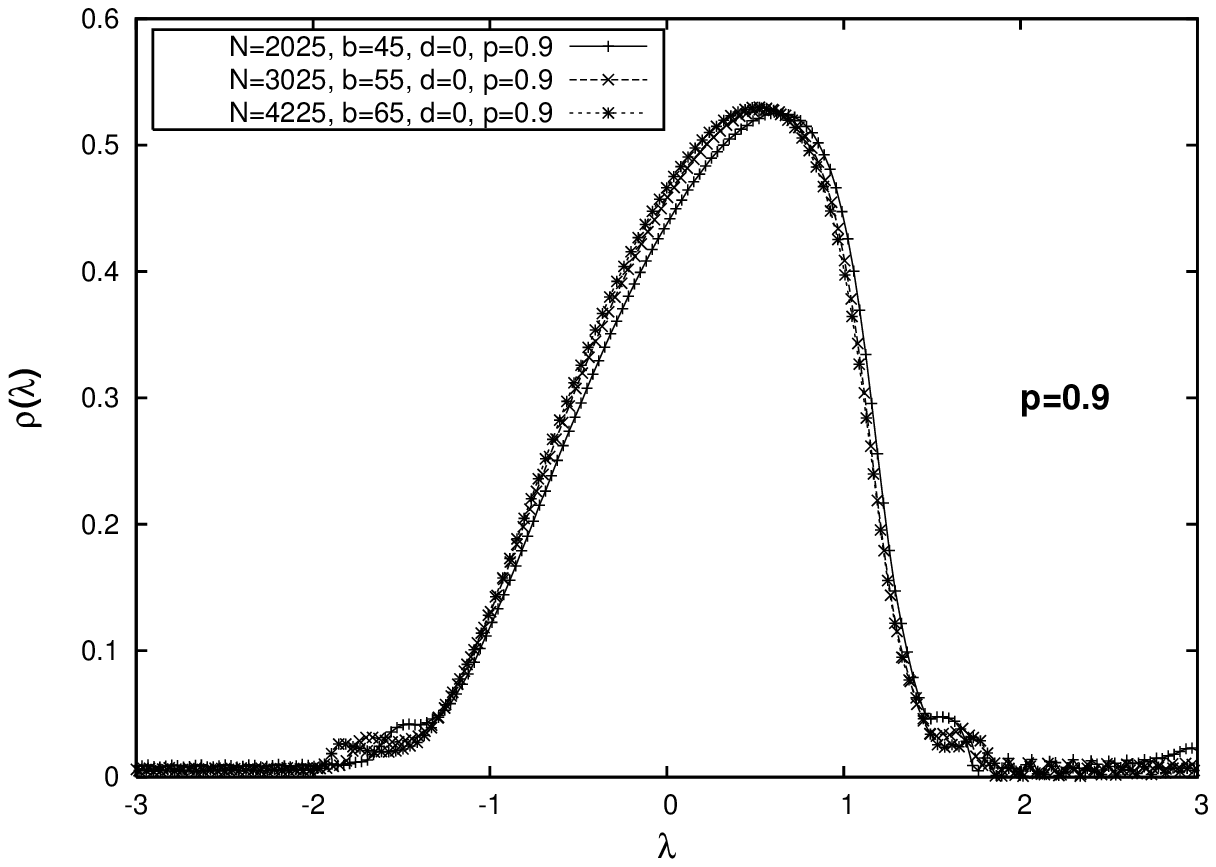}
\caption{\label{fig:p-eig}Shown is the distribution of the eigenvalues. Note that 
we have shifted the distributions by their mean values. Indicated
is also the Wigner distribution and a fit by a Gaussian.}
\end{center}
\end{figure}

Before we analyze the eigenvalue spectrum of the above defined random matrix
ensemble, we look at the structural properties of the random matrix. Since the elements of
the band are variables which can be with $-1$ or $0$ it is tempting to suspect
that the geometric distribution of these has an influence on the eigenvalues. If we view the 
random matrix with the entries $-1$ or $0$ as an adjacency matrix representing
an undirected graph (considering only the upper triangular matrix, neglecting the 
diagonal and the first off-diagonal), then the graph may be a set of unconnected 
subgraphs or may have a percolation structure. If the graph includes a percolating graph
then this introduces a divergent length scale as $N\rightarrow\infty$ possibly influencing
the eigenvalues through the correlations introduced into the main diagonal.

We define percolation in terms of a site percolation problem with free boundaries
except for the left edge entries which are always fixed to $-1$ (see above). Here, the site percolation 
probability $P_\infty$ is defined by the mass of the largest cluster divided
by the size of the band. Due to the symmetry, we only consider the upper triangular
matrix (band). Two matrix elements having both a value of $-1$ belong to the
same cluster if their indices differ at most by $1$, i.e., are nearest neighbors in the
sense of a simple square lattice~\cite{Stauffer-Aharony-1992}. 

To analyze percolation in the upper band, we use a finite-size scaling ansatz
\begin{equation}
P_\infty = b^{-\beta/\nu} {\cal P}((p-p_c)b^{1/\nu}).
\end{equation}
The finite size scaling of the percolation probability (order parameter)~\cite{Heermann1,Binder-Heermann}
for this staircased slab with aspect ratio $b^2/N =1$ is shown in Figure~\ref{fig:p-infty}. 
The critical value for the percolation threshold $p_c$ for the simple square lattice 
is $p_c = 0.592746$~\cite{Stauffer-Aharony-1992}
and the  exact values for the critical exponents  are $\alpha = -2/3, \beta = 5/36, \nu = 4/3$~\cite{Smirnov-Werner-2001}. Assuming universality we use these values for the critical exponents and find a value 
$p_c=0.61$ for the percolation threshold that gives the best fit to a scaling function ${\cal P}$. Of course,
we don't expect $p_c$ to be a universal quantity. 
Beside possible corrections to the finite-size scaling, the percolation problem posed by the
staircased slab geometry  is seen to be in the same universality class as
the two-dimensional site percolation. 

Thus, at  $p_c$ we have a correlation length which is diverging as the linear dimension
of the matrix goes to infinity. It should be noted that if $b\ll N$, then many percolating
clusters in the $b$ direction can occur~\cite{Albano-1993}. In the geometry considered
here only one percolating cluster exists~\cite{Arcangelis-1987}.
In the following we explore the consequences of this on the
eigenvalue spectrum and in particular on the distribution of the largest eigenvalue.

Before we turn our attention on
the distribution of the largest eigenvalue, we examine briefly the distribution of the
eigenvalues themselves. As suggested by the above percolation analysis we expect
the geometric percolation to influence the eigenvalue distribution. The distribution
is shown in Figure~\ref{fig:p-eig} for three values of $p$. The data shown in the
figures were obtained by averaging over $10000$ Monte Carlo samples. Beside the
matrix sizes shown, data were also produced for matrices ranging from $N=196$ up
to $N=4225$ for $b^2/N=1,5,10$ and $d=0,10$.

First we note that for different combinations of $N$ and $b$ at fixed ratio $b^2/N=1$ 
the data scales, in agreement with other band random matrix ensembles~\cite{Casati-1991}.
As a function of $p$
the distribution undergoes a shape transition. The sign of the asymmetry changes
and the distribution is nearly symmetric close to the percolation threshold. There 
we also compare the distribution to the Wigner distribution and to a Gaussian. Neither
fits the distribution. Note that for the gaussian unitary ensemble a rapid change 
has been observed~\cite{Casati-1991,Zyckowski-1996}, as one
increases the band width, from a gaussian
to a semicircle distribution.

Let $\lambda_{\mbox{max}}(H_N)$ denote the largest eigenvalue of the random
matrix $H_N$ and
\begin{equation}
F_{N,b,d,p}(\lambda) := P_{N,b,d,p}(\lambda < \lambda_{\mbox{max}})
\end{equation}
the corresponding distribution function. For the non-banded ensembles GUE and GOE, the
distribution function is the Tracy-Widom distribution~\cite{Tracy-Widom}. This distribution 
is expected to hold for an even broader class of random matrices and the precise characterization
of the class is now being investigated~\cite{Soshnikov-1999,Baik-2005}. Known is that
for symmetric matrices with iid entries of variance $1/N$, such that all moments are
finite, the Tracy-Widom distribution holds asymptotically. If the entries decay with
a power law then it is expected that the matrices belong to a different 
universality class~\cite{Biroli-2006}.

To discover the phase transition in the behavior of the largest eigenvalue we
calculate the cumulant of the distribution of the largest eigenvalue and analyze the 
flow of the cumulant as a function of
the matrix size at fixed ratio $b^2/N = 1$ and the probability $p$ with which the matrix
elements are set. We define the fourth order cumulant as
\begin{equation}
u(N,p) \sim 1 - \langle \lambda^4 \rangle / (3  \langle \lambda^2 \rangle^2) .
\end{equation}
For a second-order phase transition the cumulant flows to three fix
points~\cite{Binder-Heermann}. Two fix points characterize the values of $p$ below and above $p_c$
and one characterizes the value of $p_c$. 

\begin{figure}
\begin{center}
\includegraphics[width=\columnwidth]{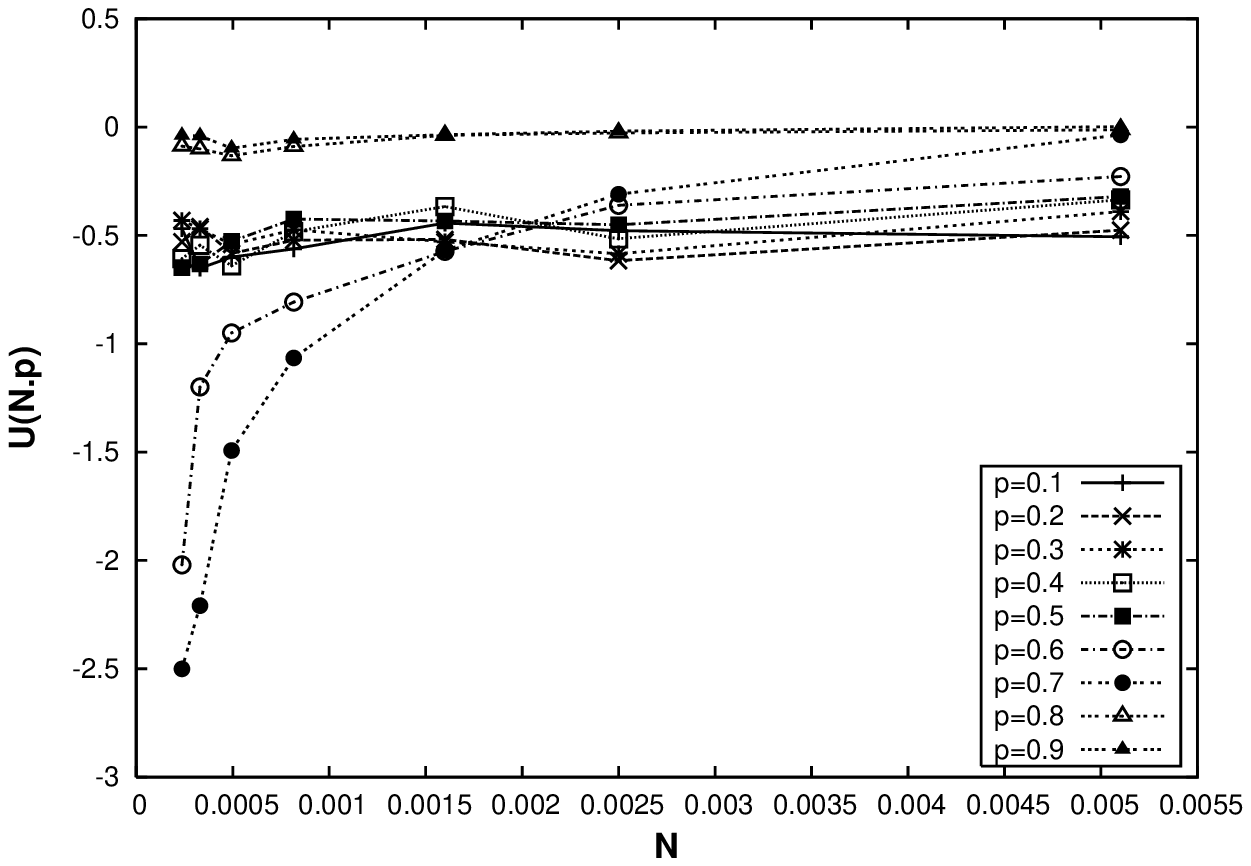}
\includegraphics[width=\columnwidth]{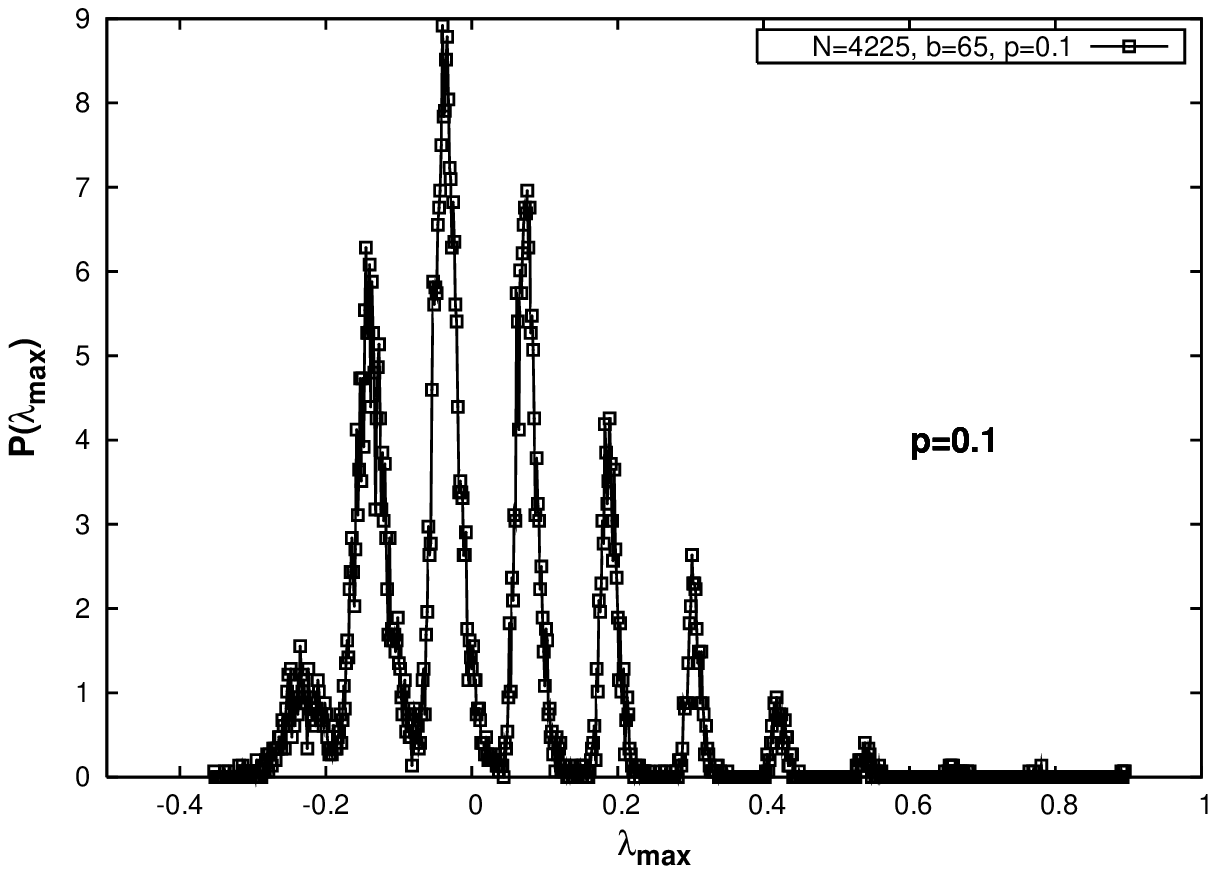}
\includegraphics[width=\columnwidth]{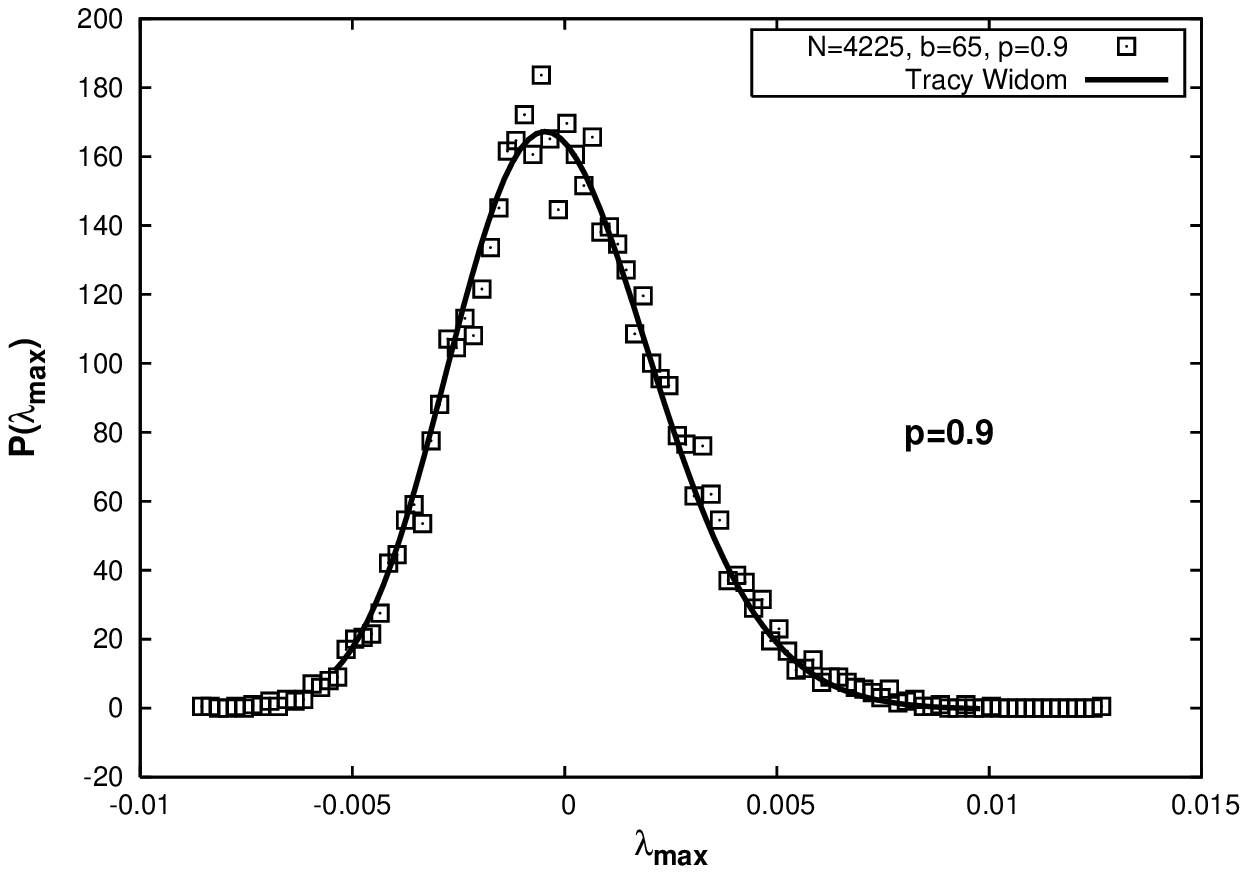}
\caption{\label{fig:u-Lp}Shown is the cumulant for the probability distribution of the largest
eigenvalue. The figure shows the cumulant $u_{N,p} = 1 - \langle(\lambda - \langle\lambda\rangle)^4\rangle / 
\langle(\lambda - \langle\lambda\rangle)^2\rangle^2$ keeping  the ratio
 $b^2/N$ fixed to $1$. Above the percolation transition at $p_c$ (roughly at $0.61$) the
flow changes and for higher values of $p$ is almost constant and zero. For values 
below $p_c$ the cumulant is fluctuating in a narrow band for the matrix sizes
considered in this work.
Also shown is the distribution at $p=0.1$ and at $p=0.9$. While below $p_c$ the
distribution is multi-peaked we find a transition to a Tracy-Widom distribution above
$p_c$.}
\end{center}
\end{figure}

In Figure~\ref{fig:u-Lp} is shown the fourth-order cumulant for the distribution of the
largest eigenvalue as well as examples of the distribution above and
below $p_c$. While the distribution of the largest eigenvalue below $p_c$ shows
multiple peaks, above $p_c$ these are reduced to a single peak. This change is
also seen in the flow of the cumulant. For values $p$ well above the percolation
transition point $p_c$ the cumulant is almost constant and nearly zero. Below the
percolation transition point the cumulant slightly fluctuates in a narrow band. Close
to the transition point the flow changes strongly indicating the transition.

In Figure~\ref{fig:u-Lp} is also shown a fit of the Tracy-Widom distribution to the
data above $p_c$ at $0.9$. The fit is very good indicating that in this ensemble
above $p_c$ the Tracy-Widom result holds. So far, for symmetric $N\times N$ matrices with
iid entries of variance $1/N$, the Tracy-Widom distribution has been shown to hold~\cite{Soshnikov-1999}, while if the 
entries fall off with a power law, the matrices fall in a in different universality class~\cite{Cizeau-1994}.
Thus it is quite surprising that for the class of random matrices defined here we do
get a Tracy-Widom distribution.

The geometric structure of a random matrix determines the
eigenvalues of a matrix to a large extend. In this paper we have introduced a class of 
band structure random matrices that pertain to the description of polymers and may 
be capable to describe the observed three-dimensional organization of chromatin. Due 
to the Bernoulli type of its off-diagonal matrix elements we are able to define a
percolation problem and link the change in the eigenvalues to the occurrence
of a percolation transition. The change is most noticeable in the density distribution
of the largest eigenvalue. The multiple peak distribution gives way to a single 
peak at the percolation transition.  But also the eigenvalue density shows a
change in skew at the transition point. Unclear is yet the type of distribution for
the eigenvalues as well as for the largest eigenvalue below the transition.

\begin{acknowledgments}
M. Bohn gratefully acknowledges  funding from the Landesgraduiertenf\"orderung. 
We are also very grateful for discussions with F. Wegner, D. Stauffer and H. Kohler.
\end{acknowledgments}



\end{document}